\documentclass[12pt]{article}
\pdfoutput=1
 \newcommand{\be}[1]{\begin{equation}\label{#1}}
 \newcommand{\ba}[1]{\begin{eqnarray}\label{#1}}
 \newcommand{\ep}[1]{\epsilon_{#1}}
 
 \newcommand{\de}[1]{\delta_{#1}}

 \newcommand{\rd}{{\rm d}}
 
 \newcommand{\re}{{\rm e}}
 \newcommand{\pa}[1]{\left(#1\right)}
 \newcommand{\paq}[1]{\left[#1\right]}

 \newcommand{\M}{{\rm M_{\rm P}}}

 \def\ee{\end{equation}}
 \def\ea{\end{eqnarray}}

 \def\mR{\mathcal R}

\usepackage{amsmath}

\usepackage{graphicx,latexsym}
\usepackage{graphicx,epsf}
\usepackage{color}
\usepackage{epsfig}
\usepackage{amsmath}
\usepackage[T1]{fontenc}
\usepackage[utf8]{inputenc}
\usepackage{tikz}
\usepackage{authblk}
\begin{document}
\title{DBI Inflation and Warped Black Holes}
\author[1]{Alexander Y. Kamenshchik\thanks{Alexander.Kamenshchik@bo.infn.it}}
\author[2]{Alessandro Tronconi\thanks{Alessandro.Tronconi@bo.infn.it}}
\author[2]{Giovanni Venturi\thanks{Giovanni.Venturi@bo.infn.it}}
\affil[1]{Dipartimento di Fisica e Astronomia, Universit\'a di Bologna and INFN, Via Irnerio 46,40126 Bologna,
Italy\\

L.D. Landau Institute for Theoretical Physics of the Russian
Academy of Sciences, 119334 Moscow, Russia}
\affil[2]{Dipartimento di Fisica e Astronomia, Universit\'a di Bologna and INFN, Via Irnerio 46, 40126 Bologna,
Italy}

\date{}

\maketitle

\begin{abstract}
We study a possible amplification mechanism for the curvature perturbations generated during inflation by a Dirac-Born-Infeld (DBI) inflaton in the presence of a sharp feature in the warp factor. The large growth resulting in the scalar power spectrum is a consequence of the decreasing speed of sound. We obtain analytical approximate expressions for the relevant dynamical quantities both in the slow roll phase and during the transient phase which leads to the amplification. We finally compare our approximations to the exact numerical evolution and give, in such a context, the general rules for building a viable inflationary model leading to, during the subsequent radiation dominated era, an abundant primordial black holes formation.
\end{abstract}

\section{Introduction}
From the very first LIGO detection of a gravitational wave signal in 2015 \cite{LIGO}, the question of whether the black holes which produced such a signal could have a primordial origin was raised. The possibility that primordial black holes (PBHs) indeed exist was investigated many years ago in the seminal papers by Zeldovich and Novikov, and by Hawking \cite{oldPBH}. At that time cosmology and astrophysics began to deal with the Dark Matter (DM) problem and PBHs soon turned out to be studied as possible DM candidates. Currently DM is definitely an urgent problem needing a solution since evidence for it is given by several independent observations. Moreover it is now clear that the primordial universe underwent a phase of accelerated expansion called inflation \cite{inflation} generating the seeds of the large scale structures and the anisotropies in the cosmic microwave background (CMB) \cite{pert}. In such a context, it is therefore tempting to investigate whether the existence of PBHs offers a further possibility of improving our comprehension of the evolution of the early universe by relating the mechanisms necessary for PBHs formation to current observations. If PBHs were formed from a gravitational collapse of some over-densities during the radiation dominated era \cite{BHform}, and such over-densities were generated during inflation, then the mass of PBHs on their formation can be related to the features of the inflaton potential in an interval which is not probed by CMB \cite{USR}. {This therefore may at least select, among a plethora of inflationary models leading to degenerate predictions concerning the scales probed by CMB, those having a potential which also leads to PBHs formation. Still discriminating between models predicting PBHs formation and leading to similar CMB spectra without assuming any further theoretical criterion is not possible. Let us note, however, that the existence of an inflationary phase generating an amplification of curvature perturbations leading to almost $\sim 100\%$ of PBHs DM requires a very peculiar tuning of the inflaton potential and therefore is very difficult to obtain. Therefore the study of new mechanisms of amplifications in non-canonical inflationary frameworks also, is certainly of interest.}\\
The quest for a theory of quantum gravity is one of the major challenges of the last 40 years of theoretical physics. Even if quantum gravitational effects are expected to become manifest at energies above the Planck scale, one still expects that inflation is sensible to trans-planckian physics and quantum gravitational effects may then leave footprints on inflationary observables. In the present paper we investigate a possible, string theory inspired, mechanism leading to PBHs formation. String theory is widely believed to provide a fundamental description of Nature and is particularly relevant at the Planck scale, where the Standard Model and General Relativity are expected to break down.\\
Despite its formulation based on simple principles, string theory predicts an enormous set of possible low energy physical realisations, the string landscape, and we expect to live in one of these realisations where physical laws have the structure we experience. Different inflationary scenarios are considered as possible in the string theory context as the scalar fields necessary to play the role of the inflaton are ubiquitous and, for example, are associated with the structure of extra-dimensions. In particular, in the Dirac-Born-Infeld (DBI) scenario \cite{DBI} the scalar field playing the role of the inflaton parametrises the position of a 3+1 D-brane moving along a six-dimensional ``throat'' with a warped geometry in the direction of motion. The peculiarity of this scenario consists of the fact that it possesses several non-standard features, some of them putting severe constraints on its viability with respect to the CMB observables related to it and others which appear indeed promising if PBHs formation is considered.\\
In the present article we study a possible mechanism for inflation and PBHs formation in the context of DBI inflation. The paper is organised as follows. In Section II we present the model, the relevant definitions and the equations governing the homogeneous dynamics. In particular, we introduce the slow roll (SR) parameters and illustrate how the dynamics simplifies during SR. We further discuss the possible existence of inflationary solutions which depart from SR and can be used to describe the evolution in the presence of sharp features in the warp factor. Section III concerns inflationary perturbations: we present the relevant equations and we discuss how a sharp feature in the warp factor may produce an amplification of the curvature perturbations. Finally a toy model is discussed in more detail and the numerical results are presented and explained. In Section IV we illustrate our conclusions.

\section{Formalism}
Let us consider the following lagrangian density for a minimally coupled DBI inflaton
\be{lagsf}
\mathcal{L}=\sqrt{-g}\paq{\frac{1}{f}\pa{1-\sqrt{1+2fX}-fV}}\equiv \sqrt{-g} P,
\ee
where $f=f(\phi)$ is related to the warp factor along a six-dimensional ``throat'' (having dimension $\paq{f}=\paq{V}^{-1}$), $V(\phi)$ is an arbitrary potential, $X=-\frac{1}{2}(\partial \phi)^2$ and $P$ is the pressure of the scalar field fluid. In order to study the existence of inflationary solutions we consider the homogeneous part of the inflaton, i.e. $\phi=\phi(t)$ with $X=-\frac{1}{2}\dot \phi^2$ on a flat FLRW background metric, described by the line element
\be{rw}
\rd s_4^2=\rd t^2-a(t)^2\rd \vec x\cdot \rd \vec x. 
\ee
The homogeneous equations of motion are the standard Einstein equations for a perfect fluid and the corresponding continuity equation, where the pressure of the fluid is defined in (\ref{lagsf}) and its energy density is 
\be{rhodef}
\rho=2 X P_{,X}-P=\frac{1}{f}\paq{-1+\frac{1}{\sqrt{1+2fX}}+fV}.
\ee
Unlike canonical models of inflation with a scalar field having $P_{\rm can}=\frac{1}{2}(\partial \phi)^2-V$, and a speed of sound $c_s$ constant and equal to the speed of light ($c_s=1$), in this context the speed of sound is variable and given by
\be{cs2}
c_s^2=\frac{P_{,X}}{P_{,X}+2XP_{,XX}}=1+2fX=1-f\dot \phi^2
\ee
with $0<c_s\le 1$ \cite{noncan}.
Let us note that a speed of sound very close to $1$ is a sufficient condition for recovering the canonical behaviour and we observe that when $f \dot \phi^2\ll 1$ such a condition is satisfied.\\
The resulting Friedmann equation is 
\be{hub}
H^2=\frac{1}{3\M^2f}\paq{-1+\frac{1}{c_s}+fV}
\ee
and the Klein-Gordon (KG) equation for the inflaton is
\be{kg}
\ddot \phi+3Hc_s^2\dot \phi+\paq{\frac{1-c_s}{2}\frac{1+2c_s}{1+c_s}\frac{f_{,\phi}}{f}}\dot \phi^2+c_s^3V_{,\phi}=0.
\ee
\subsection{Inflationary solutions}
{In what follows the conditions necessary for the existence of inflationary solutions are discussed. The relevant homogeneous equations for the inflaton-gravity system will be rewritten in terms of the slow roll parameters $\ep{i}$ and $\de{i}$ where $\ep{0}=\ln\pa{\M/H}$, $\ep{i+1}=\ep{i,N}/\ep{i}$, $\de{0}=\ln\pa{\phi/\M}$, $\de{i+1}=\de{i,N}/\de{i}$ and $\!\!\!\!\!\phantom{A}_{,N}$ is the derivative with respect to the logarithm of the scale factor. The conditions for SR inflation ($\ep{i}\sim\de{i}\ll 1$) will be then derived and the consequences of having a speed of sound $c_s\sim 1$ or $c_s\ll 1$ will be studied. The possible existence of transient regimes during inflation, when the SR conditions are violated, is then discussed. Let us note that the discussion of inflationary solutions is certainly not complete and we shall only consider the cases relevant for the rest of the article.}\\
Inflation occurs if the slow roll parameter $\ep{1}\equiv -\dot H/H^2$ is smaller than one. From the continuity equation $\dot\rho=-3H(\rho+P)$ one easily obtains 
\be{ep1}
\ep{1}\equiv-\frac{\dot H}{H^2}=\frac{3}{2}\pa{1+\frac{P}{\rho}}=\frac{3}{2}\,\frac{1-c_s^2}{1+c_s\pa{fV-1}},
\ee
which corresponds to the second Friedmann equation, and the first Friedmann equation (\ref{hub}) can be cast in the form
\be{hub2}
\M^2 H^2\pa{3-\frac{2\ep{1}}{1+c_s}}=V.
\ee
When $\ep{1}\ll 1$ Eq. (\ref{hub2}) takes the standard slow roll (SR) form
\be{hubSR}
H^2\simeq \frac{V}{3\M^2}
\ee
independently of the value of the speed of sound $c_s$.\\
The second Friedmann equation (\ref{ep1}) can be rewritten in the more standard form
\be{fr2}
\ep{1}=\frac{\phi_{,N}^2}{2\M^2c_s},
\ee
and Eq. (\ref{fr2}) can be further {differentiated} leading to 
\be{dfr2}
\frac{\phi_{,NN}}{\phi_{,N}}\equiv {\de{1}+\de{2}}=\frac{\ep{c}+\ep{2}}{2}
\ee
where the SR parameter associated with the variation of the speed of sound $\ep{c}\equiv {c_s}_{,N}/c_s$, has been introduced. Eq. (\ref{dfr2}) is exact and explicitly shows that that slow variation of $\de{i}$ and $\ep{i}$ imply also $\ep{c}\ll 1$. Let us note that the condition $\ep{2}\equiv {\ep{1}}_{,N}/\ep{1}\ll 1$ is sufficient to guarantee a long inflationary phase and a negative value of $\ep{2}$ can violate SR still not preventing the existence of a long inflationary phase.\\
The derivative of the first Friedmann equation (\ref{hub2}) leads to
\be{dhub}
\ep{1}\frac{3\pa{1+c_s}-2\ep{1}+\ep{2}-\frac{c_s}{1+c_s}\ep{c}}{3\pa{1+c_s}-2\ep{1}}=-n_V\frac{\de{1}}{2}
\ee
where $n_V\equiv \frac{\rd \ln V}{\rd \ln \phi }$. To the leading order in the slow roll approximation, with $|\ep{i}|\ll 1$ and $|\de{i}|\ll 1$, Eq. (\ref{dhub}) simplifies to 
\be{dhubSR}
\ep{1}\simeq -n_V\frac{\de{1}}{2}\Rightarrow \ep{1}\simeq c_s\frac{\M^2}{2}\pa{\frac{V_{,\phi}}{V}}^2\equiv c_s \ep{V}
\ee
and $\ep{V}\equiv \pa{\M V_{,\phi}/V}^2/2$ is the SR parameter associated with the flatness of the inflaton potential. { Let us note that Eq. (\ref{dhubSR}) is a generalisation to non-canonical models of the well known SR condition in terms of the flatness of the inflaton potential and the potential must roughly satisfy $n_V={\mathcal O}(1)$ in order for the SR approximation to hold.} Let us note that (\ref{dhubSR}) takes the ``canonical'' form in the $c_s\rightarrow 1$ limit. \\
Finally, the Klein-Gordon equation can also be written in terms of the SR parameters as
\be{kg2}
\de{1}\frac{\paq{3c_s^2-\ep{1}+\de{1}+\de{2}+\frac{1-c_s}{2}\frac{1+2c_s}{1+c_s}\de{1}n_f}}{3-\frac{2\ep{1}}{\pa{1+c_s}}}=-\frac{\M^2}{\phi^2}c_s^3n_V
\ee
where $n_f\equiv \frac{\rd \ln f}{\rd \ln \phi }$.\\
On now using (\ref{dfr2}) and the derivative of the definition (\ref{cs2}) one finds
\be{epcs}
\ep{c}=\frac{c_s^2-1}{c_s^2+1}\pa{n_f\de{1}+\ep{2}-2\ep{1}}
\ee
{which is an exact equation and shows that one also needs $n_f={\mathcal O}(1)$ during SR in order for the SR parameters to be of the same order of magnitude}. Finally, { on using (\ref{dfr2}) and (\ref{epcs})}, Eq. (\ref{kg2}) simplifies to 
\be{kg3}
\epsilon_1\frac{
\left[1+\frac{\epsilon_2-2\epsilon_1}{3(1+c_s)}-\frac{c_s\epsilon_c}{3(1+\epsilon_s)^2}\right]^2}{\left[1-\frac{2\epsilon_1}{3(1+c_s)}\right]^2}=c_s\epsilon_V
\ee
and we observe that (\ref{kg3}) and (\ref{dhubSR}) coincide to the leading order in the SR approximation. { Let us note that the set of equations obtained are exact except for (\ref{hubSR}) and (\ref{dhubSR}) which are obtained in the SR approximation ($\ep{i}\sim \de{i}\ll 1$) without any further assumption on the value of $c_s$. In particular Eq. (\ref{dhubSR}) is obtained from (\ref{kg}) when $\ddot \phi\ll 3Hc_s^2\dot\phi$ and $\dot \phi^2 f_\phi/f\ll 3Hc_s^2\dot\phi$.\\
 Let us now discuss the consequences of particular values of $c_s$ on the SR approximation.} From the definition of the speed of sound (\ref{cs2}) one has
\be{cseq}
0=c_s^2+fH^2\phi_N^2-1=c_s^2+2\frac{c_sfV}{3-\frac{2\ep{1}}{1+c_s}}\ep{1}-1
\ee
and during slow roll inflation, with $\ep{1}\simeq c_s \ep{V}\ll 1$, Eq. (\ref{cseq}) becomes
\be{cseqinf}
c_s^2\simeq\frac{1}{1+\frac{2fV\ep{V}}{3}}.
\ee
Therefore if $c_s\ep{V}\ll 1$ and $fV\ep{V}\ll 1$ the speed of sound is close to unity and
\be{cscan}
c_s\simeq1-\frac{fV\ep{V}}{3}.
\ee
In contrast if $c_s\ep{V}\ll 1$ and $fV\ep{V}\gg 1$, the speed of sound is close to zero and
\be{solnoncan}
c_s\simeq \sqrt{\frac{3}{2fV\ep{V}}}
\ee
and we now observe that $fV\gg \ep{V}^{-1}$ must also hold for consistency. On now squaring each side of Eq. (\ref{solnoncan}) and using (\ref{dhubSR}) one also finds 
\be{csnoncan}
c_s \ep{1}\simeq \frac{3}{2fV}
\ee
and the more stringent condition $fV\gg \ep{1}^{-1}$ must hold in order for SR to occur when $c_s\ll 1$. Since $\ep{1}$ is smaller than unity during inflation the product $fV$ must therefore be very large.\\
{ The conditions derived above from the SR relations will be used in what follows in order to build a model of inflation with a ``canonical'' SR phase (with $c_s\sim 1$) and a ``non-canonical'' SR phase (with $c_s\ll 1$)}.\\
It can be easily verified by comparing (\ref{cseqinf}) with the definition (\ref{cs2}) that 
\be{SRnoncan}
\dot \phi^2\simeq \frac{2 V\ep{V}}{3+2fV\ep{V}}
\ee 
is a first-order non-linear dynamical equation for $\phi$ valid in the SR regime which can be solved as usual. When $fV\ep{V}\gg1$ then (\ref{SRnoncan}) simplifies to $\dot \phi^2\simeq 1/f$.\\ 
{Let us finally discuss the existence of inflationary solutions not satisfying the SR conditions.
We note that the approximate solution $\dot \phi^2\simeq 1/f$ also exists in the $c_s\rightarrow 0$ limit and, in contrast to the SR approximation, when $\ddot \phi\gg 3Hc_s^2\dot\phi$ and $\ddot \phi\gg c_s^3V_{,\phi}$.} For such a case the KG equation simplifies to
\be{kgA}
\ddot \phi+\frac{1}{2}\frac{f_{,\phi}}{f}\dot \phi^2\simeq0.
\ee
In more detail, we observe that, even if the homogeneous evolution departs from slow roll, approximate inflationary solutions can be found. In particular, on considering a transient phase with a fast variation of $f$, and thus $n_f\gg 1$, the leading contributions in the KG equation are
\be{kgtrans}
\ddot \phi+\paq{\frac{1-c_s}{2}\frac{1+2c_s}{1+c_s}\frac{f_{,\phi}}{f}}\dot \phi^2=0
\ee
with the corresponding solution
\be{transol}
c_s= \frac{1}{1+A_s f(\phi)},
\ee
where $A_s$ is an integration constant. If this transient phase follows a slowly rolling phase then, on comparing (\ref{transol}) and (\ref{cscan}) one finds
\be{As}
A_s\simeq \frac{\ep{V,on}V_{on}}{3},
\ee
where the subscript ``$on$'' indicates that the right hand side must be evaluated at the onset of the transient phase. { The product $A_s f_{on}\simeq \ep{V,on}V_{on}f_{on}/3$ is less than $1$ when the system departs from the canonical regime.} Correspondingly one may estimate the relevant SR parameters
\be{detran}
\de{1}=s_{\dot \phi}\sqrt{\frac{3}{fV}}\frac{\M}{\phi}\sqrt{1-\frac{1}{\pa{1+A_s f}^2}}
\ee
\be{eptran}
\ep{1}=\frac{3A_s}{2V}\frac{2+A_s f}{1+A_s f},\;\ep{2}=-\de{1}\pa{n_V+\frac{n_f}{2+A_s f}},\; \ep{c}=-\de{1}n_f,
\ee
where $s_{\dot\phi}$ is the sign of $\dot \phi$. { Typically, for an increasing potential, the field slowly decrease ($s_{\dot \phi}<0$), $\ep{V,t}>0$ and $A_s>0$. Then $\phi>0$, $\de{1}<0$, $\ep{1}>0$ and the signs of $\ep{2}$ and $\ep{c}$ depend on $n_f$. During a transient phase from a ``canonical'' toward a ``non-canonical'' phase $\ep{c}<0$ and $\ep{2}<0$. Let us note that $\ep{1}$ remains close to $\mathcal{O}(\ep{V,t})$ and $|\de{1}|$ decreases as $(fV)^{-1/2}$.\\
Finally we observe that Eqs. (\ref{kgtrans}) and in particular (\ref{kgA}) resemble the inflaton equation for constant roll (CR) inflation $\ddot \phi+\beta \pa{a_{,\phi}/a}\dot \phi^2\simeq 0$. The friction term, proportional to the Hubble parameter, is negligible and the warp factor variation, giving the leading contribution to the ``friction'', can be also negative. Let us note that CR inflation may have stable/unstable solutions depending on $\beta$ and may be exploited to generate an amplification of curvature perturbations during inflation in GR with a minimally coupled inflaton. In ``modified'' inflaton-gravity systems, such as this, the condition for the amplification of the curvature perturbation may be drastically changed.}
\section{Curvature Perturbations}
In the above context, the equation for the scalar perturbations can be cast in the standard form \cite{noncan}
\be{MS}
v''_k+\pa{c_s^2k^2-\frac{z''}{z}}v_k=0,
\ee
where $v_k$ is the Fourier transformed Mukhanov-Sasaki field, the prime denotes the derivative with respect to conformal time $\eta$ and $a\,\rd \eta=\rd t$. The time dependent function $z$ is defined as
\be{zdef}
z\equiv \frac{a\sqrt{\rho+P}}{c_s H}=\frac{a\M\sqrt{2 \ep{1}}}{c_s}
\ee
and 
\be{ddz}
\frac{z''}{z}=a^2H^2\paq{2-\ep{1}+\pa{3-\ep{1}}\pa{\frac{\ep{2}}{2}-\ep{c}}+\pa{\frac{\ep{2}}{2}-\ep{c}}^2+\frac{\rd}{\rd N}\pa{\frac{\ep{2}}{2}-\ep{c}}},
\ee
where the $\ep{i}$'s are small during SR inflation but may grow large (apart from $\ep{1}$) when inflation occurs far from SR (for example during some transient phase with $n_f$ large). 
The corresponding curvature perturbations are described by $\mR_k(\eta)=v_k/z$ and satisfy the equation
\be{Req}
\mR_k''+2aH\pa{1+\frac{\ep{2}}{2}-\ep{c}}\mR_k'+c_s^2k^2 \mR_k=0.
\ee
For modes outside the horizon (roughly for $c_sk/(aH)\ll 1$) the solution of (\ref{Req}) consists of a constant solution plus a varying solution of the form
\be{Rlong}
\mR_k=A_0+A_1\int^{\eta}\frac{\rd \eta'}{z^2},
\ee
where the second term in Eq. (\ref{Rlong}) is either decreasing or increasing depending on the sign of $z'/z$. Indeed, during slow roll inflation the parameters $\ep{2}$ and $\ep{c}$ in the friction term are much smaller than one and $\mR_k$ freezes at the horizon exit. In contrast if $\ep{2}/2-\ep{c}<-1$, the friction term has a negative sign and $\mR_k$ may grow after the horizon exit. An inflationary phase with a negative friction term in (\ref{Req}) is often exploited to build models of inflation wherein a certain interval of modes of the curvature perturbations are amplified. The corresponding over-densities, which form when inflation ends, may then collapse and originate primordial black holes. For example, in canonical models of inflation ($\ep{c}=0$) the presence of an inflection point in the potential \cite{USR} leads to a phase of ultra slow roll with $\ep{2}\simeq -6$ wherein the amplitude of curvature perturbation is amplified. In non-canonical models of inflation a rapid increase of the speed of sound may be responsible for the amplification \cite{tasinato}. Still some care is due because a very fast growth of $c_s$ may delay or even prevent modes from exiting the horizon and therefore stop amplification. In particular one needs a decreasing $\frac{kc_s}{a H}$ ratio, i.e. 
\be{csgrow}
\pa{\frac{kc_s}{a H}}^{-1}\frac{\rd}{\rd N}\frac{kc_s}{a H}<0\Rightarrow \ep{c}<1-\ep{1}\;{\rm and}\;\ep{c}>1+\frac{\ep{2}}{2}
\ee
must be satisfied and this necessarily requires a careful tuning of the inflationary model.\\
The resulting power spectrum is given by the expression
\be{defPS}
{\mathcal P}_\mR=\left.\frac{k^3}{2\pi^2}\left| \mR_k\right|^2\right|_{\rm e.o.i.}
\ee
where the subscript indicates that it must be evaluated at the end of inflation (e.o.i.).
When the SR conditions are satisfied the curvature perturbations freeze after the horizon exit. In such a case their power spectrum can be evaluated at the horizon exit (h.e.) and is given by
\be{PS}
{\mathcal P}_\mR=\left.\frac{1}{8\pi^2\M^2}\frac{H^2}{c_s\ep{1}}\right|_{\rm h.e.}\simeq \left.\frac{V}{24\pi^2\M^4}\frac{1+\frac{2}{3}fV\ep{V}}{\ep{V}}\right|_{\rm h.e.}
\ee
and thus an amplification or suppression of the spectrum can be realised as a consequence of a corresponding suppression or amplification of the product $\ep{1}c_s$ or equivalently by the amplification/suppression of the product $fV$, provided the SR conditions are not violated. If SR conditions are violated but $z'/z$ is positive at horizon crossing one may still evaluate $\mathcal{P}_\mR$ at horizon exit which now occurs when $c_sk\sim aH|\gamma|$ and $\gamma\neq 1$ is implicitly defined as $z''/z=2a^2H^2\gamma^2$. The resulting expression is 
\be{PSnoSR}
{\mathcal P}_\mR=\left.\frac{1}{8\pi^2\M^2}\frac{H^2|\gamma|^2}{c_s\ep{1}}\right|_{\rm h.e.}=\left.\frac{1}{12\pi^2}\frac{V|\gamma|^2}{\de{1}^2\phi^2\M^2}\right|_{\rm h.e.}
\ee
and coincides with (\ref{PS}) when $\gamma\simeq 1$. If $|\gamma|$ is large and/or $\ep{1}c_s$ is smaller compared to its value at CMB scales the resulting amplitude gets amplified.\\
Let us note that when modes exit the horizon and freeze, they still remains constant if, at some time, $z'/z$ becomes negative because $A_1\ll 1$ in (\ref{Rlong}). In contrast, if $A_1$ is much larger (and this occurs for modes exiting the horizon when $z'/z<0$) a significant amplification of the curvature $\mR_k$ could be produced if the condition $z'/z<0$ is valid for a certain time interval (or number of e-folds $\Delta N_{\rm amp}\ge 1$).\\
A mechanism leading to an abundant amplification of $\mR_k$ was already discussed in a previous paper \cite{ktvv} in the context of non-canonical models of inflation. There the evolution of an inflationary model with a modified kinetic term plus a potential was studied. The presence of the potential is necessary in order to drive a SR phase wherein the perturbations we observe in the CMB are generated. In contrast, the modified kinetic term, which becomes leading at some point during inflation and after the SR phase, is responsible for the amplification of the scalar perturbations and the subsequent formation of PBHs. { Therefore in such a model, owing to a decreasing speed of sound, the decaying mode is present and becomes negligible soon after horizon exit. The amplitude of the spectrum however is amplified w.r.t. the ``canonical'' phase since $c_s$, in the denominator of (\ref{PSnoSR}) becomes smaller and smaller. An analogous mechanism will be considered in the present article for DBI inflation.}\\
As we already discussed, for DBI inflation the inflaton lagrangian has a SR solution at $c_s\sim 1$ and at $c_s\sim 0$ and the curvature perturbations can be therefore amplified in the transition from the first phase to the second. In order for the over-densities to collapse and form a sufficient population of PBHs with mass $m_{\rm bh}\sim 10^{-15}{\rm M_{\odot}}$ (thus possibly forming a large fraction of the Dark Matter budget today {- see for example \cite{carr} and references therein}) the amplification must increase the power spectrum until the critical threshold ${\mathcal P}_\mR\sim 10^{-2}$ is reached. The modes affected by such an amplification must exit the horizon roughly at
\be{DNamp}
\Delta N(m_{\rm bh})\sim 26-\frac{1}{12}\ln \pa{\frac{g_{*k}}{g_{*0}}}+\ln\frac{c_{s,{\rm bh}}H_*}{c_{s,*}H_{\rm bh}}
\ee
e-folds after CMB modes, where $g_{*k}$, $g_{*0}$ indicates the number of relativistic degrees of freedom at the BHs formation and at present respectively, $c_{s,*}$ and $H_*$ are evaluated at horizon exit for CMB modes and $c_{s,{\rm bh}}$ and $H_{\rm bh}$ must be evaluated when the spectrum reaches the critical threshold. { Let us note that the critical threshold $\sim 10^{-2}$ is estimated by also assuming that the curvature perturbations responsible for the collapse are gaussian (other effects are also very important for determining the threshold). Among the others, the effect of non-gaussianities on the estimate of the threshold may be quite large for models with a varying speed of sound, and the problem is still being debated. A precise estimate goes beyond the scope of this paper and would require a series of strong assumptions on the dynamics of the collapse. The relevant point we highlight in this article is to illustrate a mechanism for the amplification of the curvature perturbations and relate the resulting spectrum to the parameters of the model.}
 
\subsection{Model building}
We now illustrate how a viable DBI inflationary model which includes an amplification phase may be realised. The features of the scalar inflationary perturbations imprinted in the CMB put severe constrains on the inflaton evolution when these perturbations modes exit the horizon, $\sim60$ e-folds before inflation ends. CMB observations \cite{Planck} constrain the spectral index $n_s-1\sim {\mathcal O}(10^{-2})$, the amplitude of the spectrum ${\mathcal P}_\mR\sim 2\cdot 10^{-9}$ and the speed of sound (which can be related to the statistical abundance of non-gaussian features observed in the CMB \cite{nongaus}) 
\be{csconst}
c_s>3\cdot 10^{-2}.
\ee 
If we assume that $c_s\sim 1$ at that time the remaining two constraints to be satisfied are
\be{PSSR}
{\mathcal P}_\mR=\frac{1}{24\pi^2\M^4}\frac{V_*}{\ep{V,*}}\sim 2\cdot 10^{-9}, \quad n_s-1\sim {\mathcal O}(\ep{V,*})\sim {\mathcal O}(10^{-2}),
\ee
where the subscript $\!\!\!\!\!\phantom{A}_*$ indicates that the quantities are evaluated when CMB scales exit the horizon. This fixes the shape of the potential, i.e. its slope and value at the ``beginning'' of inflation and the magnitude of $f_*$ (as we already discussed $fV\ep{V}\ll 1$ is needed in order for $c_s$ to be very close to unity).\\
Hereafter we shall study how the amplification of the spectrum ${\mathcal P}_\mR$ may occur as a consequence of a sharp feature in the warp factor $f$. The sharpness is required in order to realise the amplification needed for an efficient PBHs production in the short amount of time of a few e-folds. The perturbations which exit the horizon $\sim 30$ e-folds after those imprinted in the CMB would collapse in PBHs of particular interest (see (\ref{DNamp})) since they may survive evaporation until today and constitute a large part of the Dark Matter budget.\\
\begin{figure}[t!]
\centering
\includegraphics[width=6.5cm]{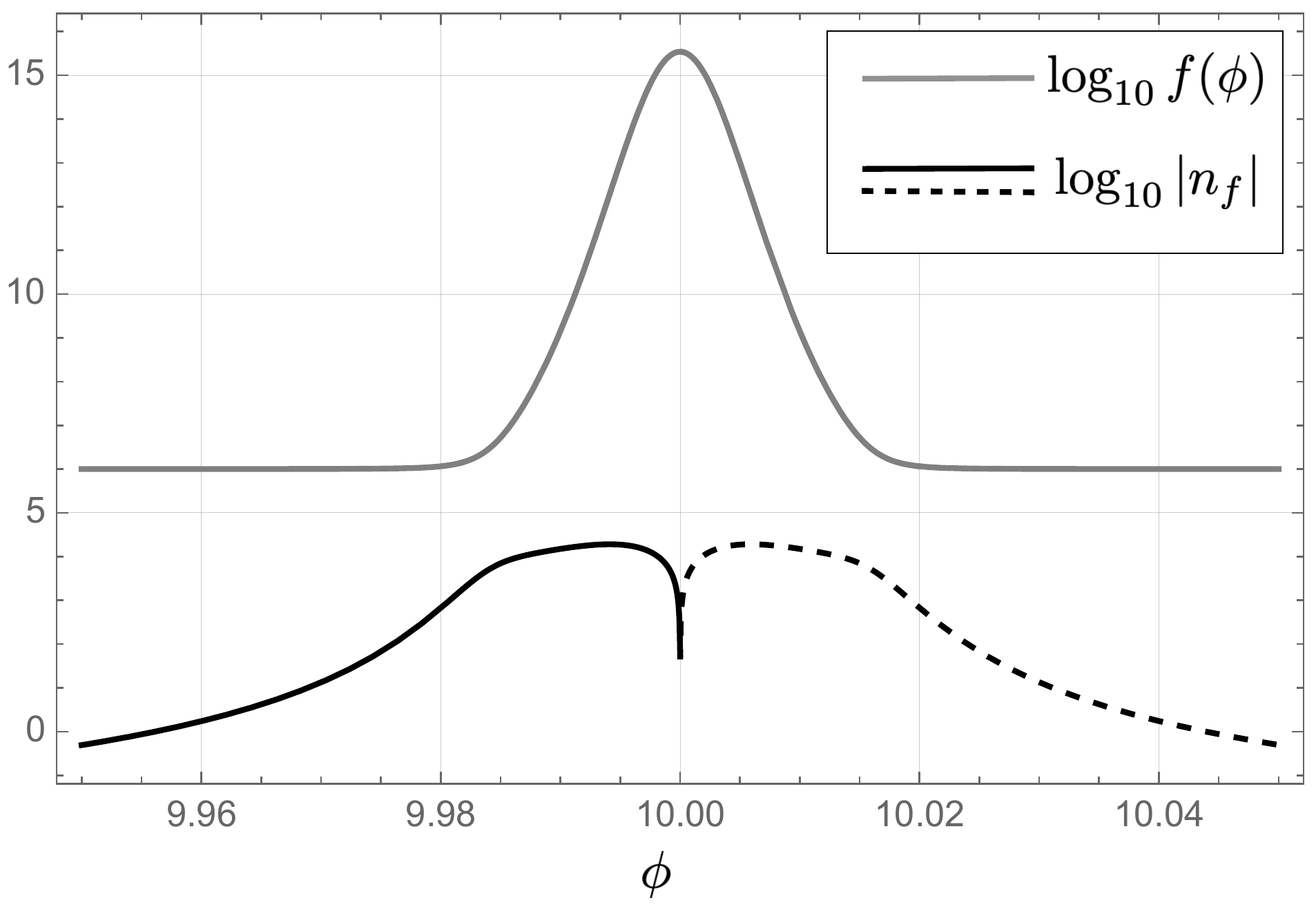}
\includegraphics[width=6.5cm]{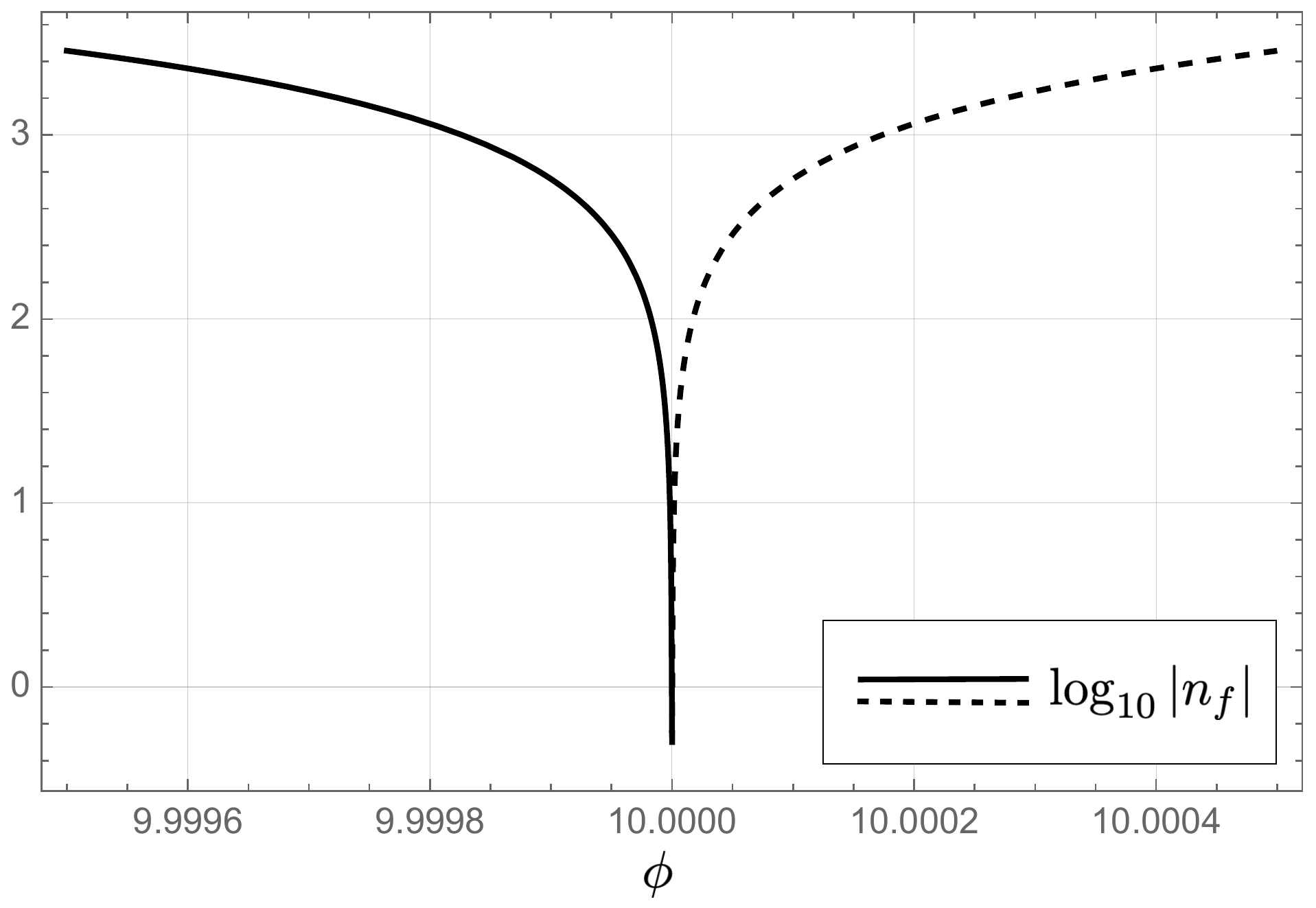}
\caption{\it In the figure on the left $\log_{10}f(\phi)$ is plotted as a solid, grey line and $\log_{10}|n_f(\phi)|$ is plotted as a solid black line for the interval where $n_f(\phi)>0$ and a dashed black line in the interval where $n_f(\phi)<0$. The plot refers to the warp factor leading to the evolution we discussed, and is described by (\ref{feat}) with (\ref{numfun}) In the figure on the right $\log_{10}|n_f(\phi)|$ is plotted in a smaller interval around the peak at $\phi_0=10\M$. In the figures we assume $\M=1$.
\label{figf}}
\end{figure}
Let us note that the possible physical origins of the presence of a sharp feature in the warp factor have been studied in several articles \cite{feature} and we shall not discuss it further. { In particular the existence of such features at scales probed by CMB has been studied, however one may consider their presence at smaller energies, not testable by CMB.} We shall limit ourselves to analysing its consequences as a possible source of amplification of the curvature perturbations. In order to simplify our model we shall study the consequences of a gaussian feature in an interval with a constant warp factor. Nonetheless one may easily generalise our results to more realistic models.\\
When the scalar field passes through the region where the feature is present the sound speed decreases and the dynamics may enter a non-canonical phase. Thus, as a consequence, the spectrum of the scalar perturbations gets amplified. One must estimate the duration of such a transient phase and tune the parameters of the model to ensure that the inflaton crosses the feature within a few e-folds.
Let therefore consider
\be{feat}
f=F\paq{1+A\re^{-\frac{\alpha}{\M^2}(\phi-\phi_0)^2}},
\ee
{ where $\alpha^{1/2}$ is inversely proportional to the width of the peak in $\M$ units, $\phi_0$ is the position of the feature of interest, $A$ is the height of the peak and $F$ is the constant value of the warp factor far away from the peak. While $n_f\sim 0$ away from the peak ($|\phi-\phi_0|\gg 1$) and in a narrow interval around the peak ($\phi\sim \phi_0$) it may grow very large while approaching to $\phi_0$. Let us note (see (\ref{cscan}) and (\ref{solnoncan})) that the product $f V\ep{V}$ determines the value of $c_s$ during SR inflation. 
In detail the evolution of the inflaton is described by the following phases:
\begin{itemize}
\item[i)] the inflaton slowly rolls toward the peak, $F V\ep{V}\ll 1$ and $c_s\simeq 1$; in this phase CMB features are generated;
\item[ii)] the inflaton encounters the gaussian feature and climbs. In such a phase if $n_f$ is large, $|\dot \phi|$ rapidly approaches the value $\sim 1/\sqrt{f}$ (see (\ref{kgA})) and further, If $F A V\ep{V}\gg 1$, the speed of sound is $c_s\simeq 0$ (see (\ref{SRnoncan}));
\item[iv)] the inflaton goes down the peak, $n_f$ is large and the speed of sound increases, $c_s\rightarrow 1$;
\item[v)] the inflaton slowly rolls down the potential until inflation ends.
\end{itemize}
}
When the inflaton evolves non-canonically then $|\dot\phi|\sim f^{-1/2}$ and 
\be{DNnc}
\Delta N_g=\int_{t_i}^{t_f} H\rd t=\int_{\phi_i}^{\phi_f}\frac{H}{\dot \phi}\rd \phi
\ee
is the number of e-folds needed to cross the (gaussian) feature. Between $\phi_i$ and $\phi_f$ the gaussian is much larger than $1$ and one has
\be{DNnc}
\Delta N_g\simeq\int_{\phi_i}^{\phi_f}\sqrt{\frac{VFA}{3\M^2}}\re^{-\frac{\alpha}{2\M^2}(\phi-\phi_0)^2}\rd \phi\sim \int_{-\infty}^{+\infty}\sqrt{\frac{V_{0}F A}{3\M^2}}\re^{-\frac{\alpha}{2\M^2}(\phi-\phi_0)^2}\rd \phi,
\ee
where in the last integral we approximated $V$ by its value at the maximum of the gaussian, $\phi_0$, and we extended the integral over the whole real axis because the contributions to the integral from the tails of the gaussian function are negligible. We then obtain the following estimate
\be{DNfin}
\Delta N_g\sim\sqrt{\frac{2\pi V_0 F A}{3\alpha}}
\ee
which can be expressed in terms of the cosmic time as
\be{Dtnc}
\Delta t_g=\int_{N_i}^{N_f}\frac{\rd N}{H}\sim \sqrt{\frac{3\M^2}{V_0}}\Delta N_g.
\ee
Furthermore, close to the maximum of the gaussian $n_f$ is $\mathcal{O}\pa{1}$ and SR is restored.\\
{Let us now evaluate the curvature perturbations spectral amplitude generated by the gaussian feature and compare such an amplitude with that, fixed by observations, at CMB scales. The amplitude of the spectrum generated around $\phi_0$, when the inflaton slowly rolls, is given by (\ref{PS}), with $fV\ep{V}\gg 1$, and takes the form  
\be{psamp}
\mathcal{P}_{\mR,0} =\frac{V_0}{24\pi^2\M^4}\frac{2f_0V_0}{3}={\mathcal P}_\mR\frac{2f_0V_0^2\ep{V,*}}{3V_*},
\ee
where the last equality contains the amplitude ${\mathcal P}_\mR$ at the CMB scales.
In order for ${\mathcal P}_{\mR,0}$ to be amplified by $n$ orders of magnitude w.r.t. ${\mathcal P}_\mR$ one then needs
\be{ampcond}
\frac{\mathcal{P}_{\mR,0}}{\mathcal{P}_{\mR}}\simeq\frac{2f_0V_0^2\ep{V,*}}{3V_*}\sim 10^n.
\ee
One may also evaluate the spectrum generated by the transient phase before $\phi_0$ is crossed, which still occurs for $\phi\sim \phi_0$ if the gaussian is sharply peaked around $\phi_0$. A rough estimate of (\ref{PSnoSR}) on using Eqs. (\ref{detran},\ref{eptran}) leads to $|\gamma|^2\sim {\mathcal O}(n_f^2\de{1}^2)$ and
\be{PSnoTR}
{\mathcal P}_{\mR,t}\sim\left.\frac{1}{12\pi^2}\frac{V_0n_f^2}{\phi_0^2\M^2}\right|_{\rm h.e.}\sim\left.\frac{1}{12\pi^2}\frac{\alpha V_0}{\M^4}\right|_{\rm h.e.}= \frac{2\alpha V_0\ep{V,*}}{V_*}{\mathcal P}_{\mR},
\ee
when $n_f\gg 1$ and with ${\rm max} \;\left|n_f\right|\sim{\mathcal O}\pa{\phi_0\M^{-1}\sqrt{\alpha}}$. In order for ${\mathcal P}_{\mR,t}$ to be amplified of $n$ orders of magnitude w.r.t. ${\mathcal P}_\mR$  one now needs
\be{ampcondtr}
\frac{\mathcal{P}_{\mR,t}}{\mathcal{P}_{\mR}}\simeq\frac{2\alpha V_0\ep{V,*}}{V_*}\sim 10^n.
\ee
In the following sections, using the above expressions, we shall investigate the consequences of the particular choice of the warp factor (\ref{feat}), which for the sake of clarity is plotted in the figure (\ref{figf}), and of the potential.}
\subsection{Numerical Example}
We finally illustrate a simple applications of our estimates from the previous section and then compare the resulting curvature perturbations spectrum to the numerical (exact) results. For simplicity we consider an exponential potential driving SR inflation
\be{pot}
V=V_*\exp\paq{\frac{\beta\pa{\phi-\phi_*}}{\M}}
\ee
with $\beta=3\cdot 10^{-1}$, a constant $\ep{V}=4.5\cdot 10^{-2}$ and a warp factor given by the expression (\ref{feat}). The amplitude (\ref{PSSR}) fixes 
\be{Vstar}
V_*\sim 2\cdot 10^{-8}\M^4
\ee
and $F$ should be chosen so as to ensure that the sound speed satisfies the experimental constraints. The sound speed must be close to one when CMB scales exit the horizon and therefore we set $F=10^{6}\,\M^{-4}$. With such a setup the field slowly rolls down the potential. If we assume that amplification starts $\Delta N_*\sim 30$ e-folds after CMB scales exit the horizon one has 
\be{DNDphi}
\phi_0\simeq \phi_*-\beta^2 \M\Delta N_*.
\ee
One may therefore estimate (\ref{DNfin})
\be{DNfinest}
\Delta N_g\sim\sqrt{\frac{2\pi F V_*A}{3\alpha}}\exp\paq{-\frac{\beta^2\Delta N_*}{2}}
\ee
and the amount of amplification realised during the transient can be related with the parameter $\alpha$ in the gaussian as follows {(see (\ref{ampcondtr}))}
\be{transamp}
\frac{\mathcal P_{\mR,t}}{\mathcal P_{\mR}}=2\alpha \re^{-\beta^2\Delta N_*}\ep{V,*}\sim 10^n\Rightarrow \alpha\sim 10^{n+2}.
\ee
Near the maximum of the warp factor $f$, SR is restored and the amount of amplification is the following function of the product $F\cdot A$ {(see (\ref{ampcond}))}
\be{SRamp}
\frac{\mathcal P_{\mR,0}}{\mathcal P_{\mR}}=\frac{2}{3}FA V_*\re^{-2\beta^2 \Delta N_*}\ep{V,*}\sim 10^n\Rightarrow F \cdot A\sim 3\cdot 10^{n+12}.
\ee
Let us note that by comparing the ratio of (\ref{SRamp}) and (\ref{transamp}) with (\ref{DNfinest}) one has	
\be{ratioamp}
\frac{\mathcal P_{\mR,0}}{\mathcal P_{\mR,t}}=\frac{FAV_*\re^{-\beta^2 \Delta N_*}}{3\alpha}=\frac{\Delta N_g^2}{2\pi}
\ee
and we therefore observe that for a transient phase lasting ($\sim \Delta N_g/2$) few e-folds the two expressions lead to the same amount of amplification.\\
\begin{figure}[t!]
\centering
\includegraphics[width=9.5cm]{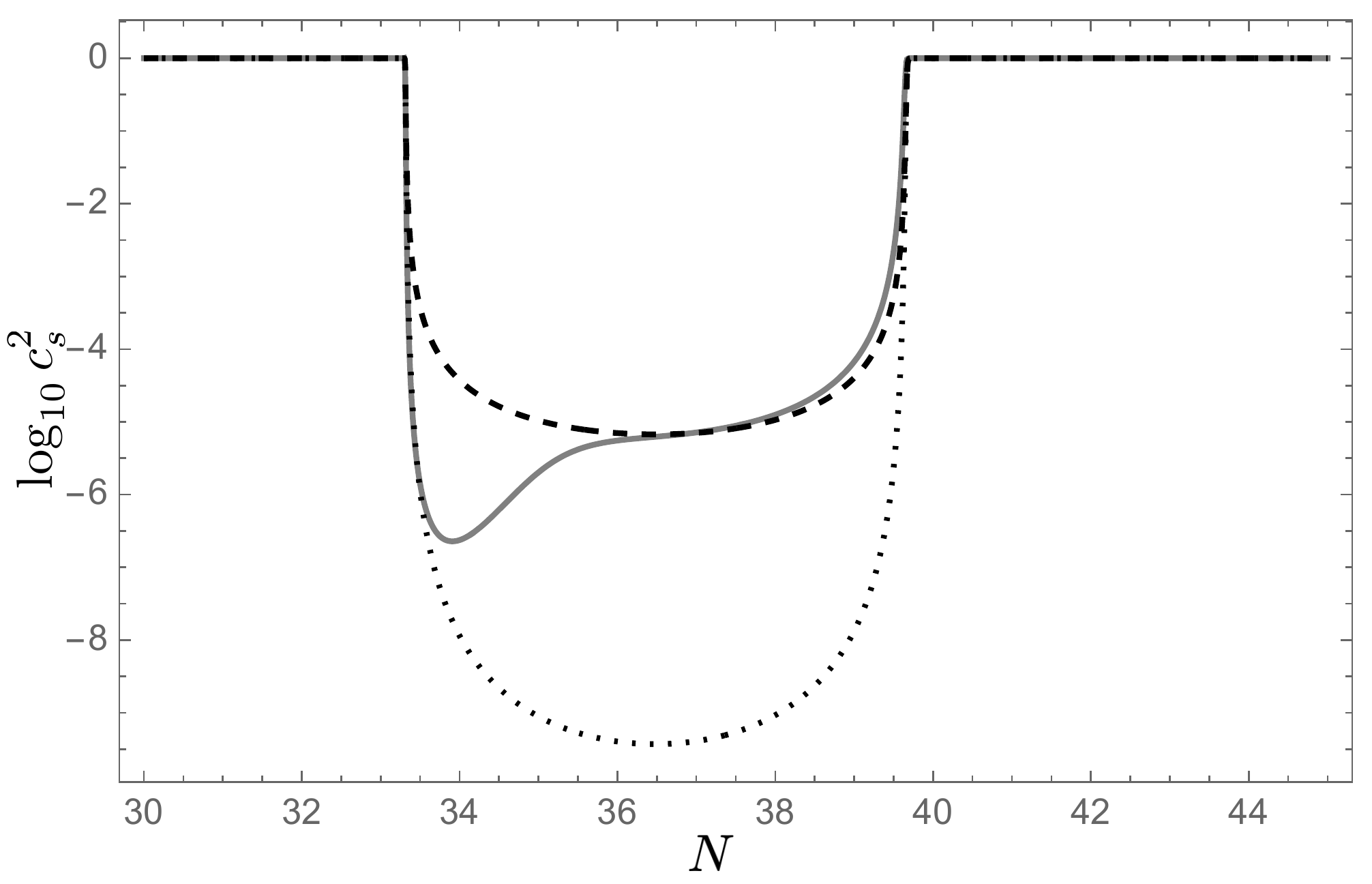}
\caption{\it In the figure above the evolution of $\log_{10}c_s^2$ is plotted as a function of $N$. The solid grey line represents the exact numerical solution of the homogeneous equations, the dashed line represents the SR analytical approximation and the dotted line represents the approximation found for the transient phase. The interval of the transient which includes its beginning and the period of maximum variation of $c_s$ is very well described. Moreover the figure shows that the estimate for $\Delta N_g$ is well approximated by (\ref{DNfinest}).
\label{fig}}
\end{figure}
\subsection{Numerical Results}
For more accurate results and in order to test the analytical approximate expressions obtained, a numerical analysis is required.\\ 
\begin{figure}[t!]
\centering
\includegraphics[width=6.5cm]{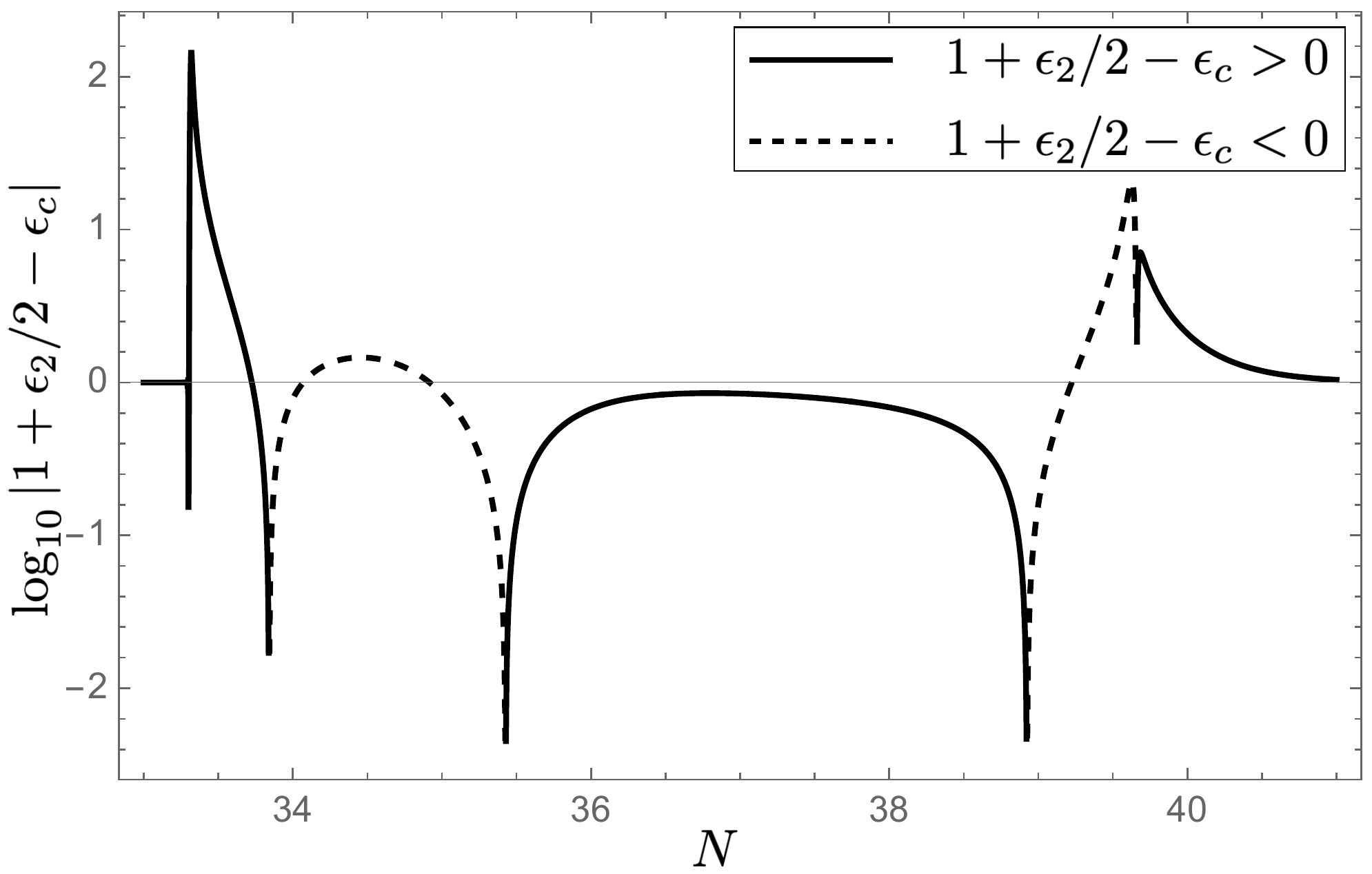}
\includegraphics[width=6.5cm]{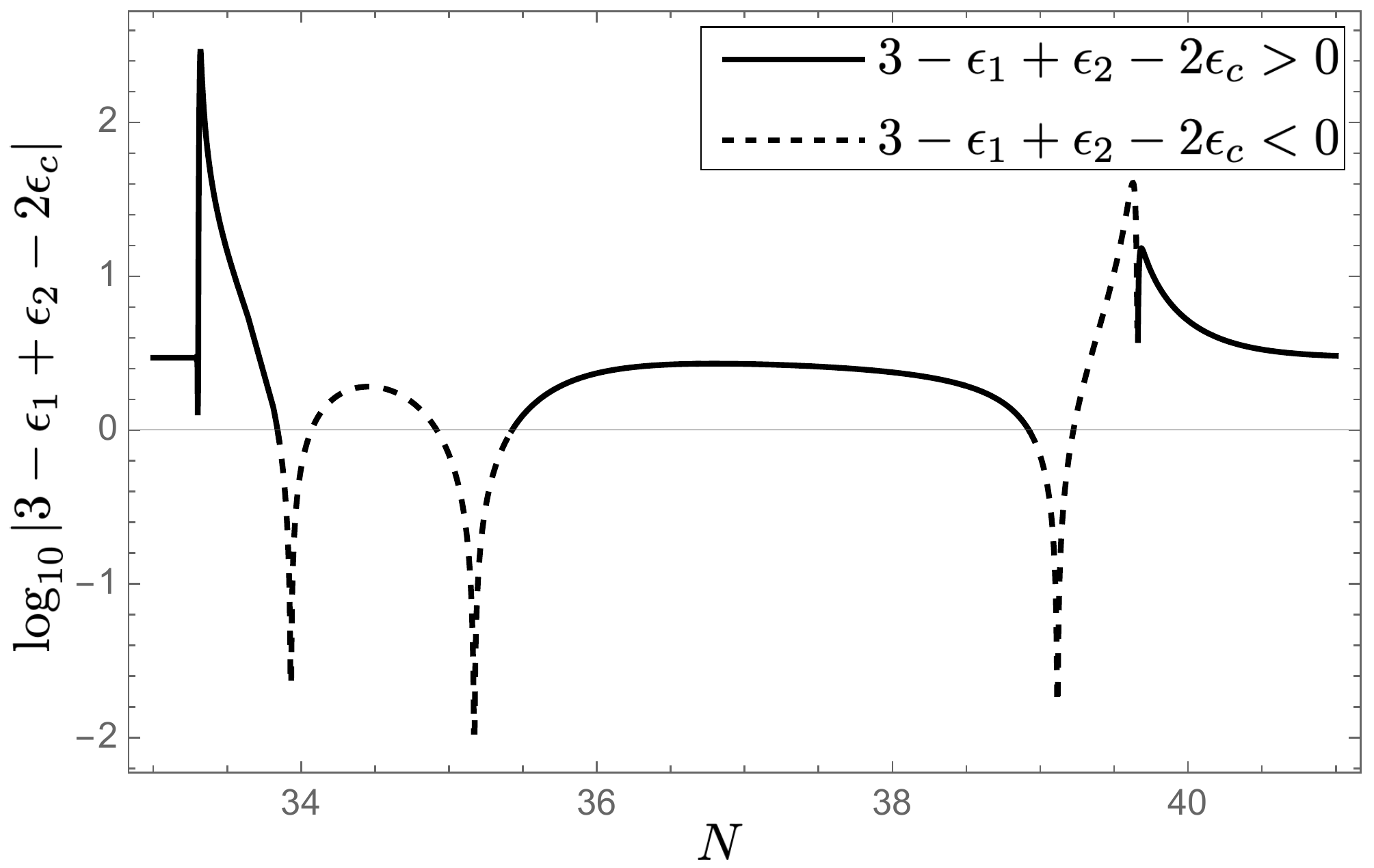}
\caption{\it In the figure on the left the behaviour of the friction term in the Eq. (\ref{Req}) is shown for the case under study and results in the spectrum plotted in (\ref{fig2}). The intervals with a dashed line are those where the friction is negative and lead to an amplification of the curvature perturbations. Let us note that the amplification phase is short (lasting a few e-folds) and correspondingly the amplifying ``friction'' is quite small (at least compared to USR and similar mechanisms of amplification). In the figure on the right the corresponding quantity $\xi$ defined by (\ref{xidef}) and (\ref{dRdN}) is plotted.  
\label{figz}}
\end{figure}
The plot in the figure (\ref{fig}) shows the homogeneous function $\log_{10}c_s^2$ as a function of $N$ in order to compare the exact (numerical) evolution of the homogeneous system (solid grey line) and the analytical approximations for the SR evolution (dashed line) and the transient (dotted line). The approximations are very close to the exact expression. Let us note that $c_s^2-1$ is proportional to $f$ which is very large in the interval when the gaussian feature is present and thus $c_s^2$ is very sensitive to very small differences between the exact and the approximate evolutions. Very small differences between the exact evolution of $\phi(N)$ and the corresponding analytical approximations are amplified by several order of magnitude when multiplied by $f$ and, as the figure shows, result in just a few order of magnitude deviations as far as the evolution of $c_s$ is concerned.\\
{ In the figure (\ref{figz}) we plotted two quantities. In the plot on the left the friction term behaviour in Eq. (\ref{Req}) as a function of $N$ is displayed. When the friction is negative (dashed line) the curvature perturbations are amplified. This occurs (roughly) in the two short intervals $N\in[34,35.5]$ and $N\in[39,39.5]$ where the friction term is quite small. The plot on the right illustrates the behaviour of the quantity
\be{xidef}
\xi\equiv 3-\ep{1}+\ep{2}-2\ep{c}
\ee
which is associated with the variation of $\mR_k$ by 
\be{dRdN}
\frac{\rd \mR_k}{\rd N}=C_k\exp\paq{-\int\xi \,\rd N}.
\ee
If SR conditions are satisfied $\xi\sim 3>0$ and $\rd \mR_k/\rd N\rightarrow 0$ in a few e-folds. Correspondingly $\mR_k$ freezes. 
In contrast, in the intervals $N\in[34,35.5]$ and $N\in[39,39.5]$, $\xi$ becomes negative and slightly delays the freezing of $\mR_k$.
Let us note that, while the friction term and $\xi$ become negative for a short amount of time during inflation due to the presence of the gaussian feature in the warp factor, their effect on the total amplification of $\mR_k$ is still negligible in comparison to that which is originated by the decrease of $c_s$ and $\ep{1}$. This fact is confirmed by the numerical estimate of the spectrum on which we shall comment below.} 
In the figure (\ref{fig2}) we plotted the spectrum $\log_{10}{\mathcal P}_{\mR}$ as a function of $\log_{10}k/a_0$ and $a_0$ is the arbitrary value of the scale factor at the beginning of the numerical analysis. The solid line represents the exact numerical results while the dots corresponds to the analytical estimate (\ref{PSnoSR}) evaluated at horizon crossing, where the horizon exit is estimated for each mode by solving exactly the condition $c_s k=aH|\gamma|$ through the (homogeneous) exact numerical solutions. 
We set 
\be{numfun}
\alpha\simeq 2.9\cdot 10^5,\;F=10^6,\; F\,A\simeq 3.5 \cdot 10^{15},
\ee
which, according to our estimates, gives an amplification phase of about $\Delta N_0\sim 4$ e-folds (it takes $\sim 8$ e-folds for the scalar field to get across the gaussian feature) and an amplification factor of about $3$ orders of magnitude.\\
Let us note that (\ref{PSnoSR}) well approximates the numerical results before and after the transient stage and when the field enters in the region where $|n_f|$ becomes large ($c_s$ decreases) at the onset of the gaussian feature. Close to the maximum of $f$ the homogeneous evolution enters a short SR phase and finally $|n_f|$ becomes large again ($c_s$ increases) as the field crosses the maximum of the gaussian feature and finally relaxes on the SR attractor with $f={\rm const}$. The figure shows that after the first large variation of $f$ the amplitude ${\mathcal P}_{\mR}$ reaches a maximum and then begins oscillating with an increasing frequency. This oscillation is the consequence of modes re-entering the horizon due to the increasing speed of sound. At the horizon re-entry $v_k$ evolves as a superposition of ingoing and outgoing waves 
\be{v_k}
v_k\sim \frac{1}{\sqrt{2c_sk}}\paq{C_k\re^{i\int_{\eta_k}^\eta kc_s\rd \eta'}+D_k\re^{-i\int_{\eta_k}^{\eta} kc_s\rd \eta'}},
\ee
where $\eta_k$ is the conformal time at re-entry, and therefore its modulus oscillates. Let us note that at early times, before exiting the horizon for the first time, the Bunch-Davies initial condition sets $D_k\rightarrow 0$ for each mode and modulus oscillations are absent. At the end of inflation, the amplitude of the modes outside the horizon is frozen and the oscillation remains imprinted in the perturbations.\\
{ In the figure (\ref{figosc}) we plotted the two contributions in the time dependent frequency of the Mukhanov-Sasaki equation (\ref{MS}) rescaled by $a^2H^2$. When $k^2c_s^2\gg |z''/z|$ the modes are ``inside'' the horizon and oscillate; in contrast when $k^2c_s^2\ll |z''/z|$ the modes are ``outside'' the horizon and the curvature perturbations evolve according to (\ref{Rlong}). If $k\,c_s$ slowly evolves monotonically and $z''/z$ rapidly  increases monotonically (as in common inflationary models) the two regimes are sequential for each mode. For the case under study, as Fig. (\ref{figosc}) shows, a certain bunch of perturbation modes may cross the horizon several times (see for example the dashed line which finally ``exits the horizon'' at $N\sim 41$). The resulting effect for this bunch of modes is that they are in a superposition of the form (\ref{v_k}) before the last ``horizon exit'' and an oscillatory region of the spectrum is generated. Let us note that the oscillation is more evident if $|C_k|\sim|D_k|$ in (\ref{v_k}). The modes which do not belong to the above subset just cross the horizon once and the resulting part of spectrum shows no evident oscillation.}\\
\begin{figure}[t!]
\centering
\includegraphics[width=9.5cm]{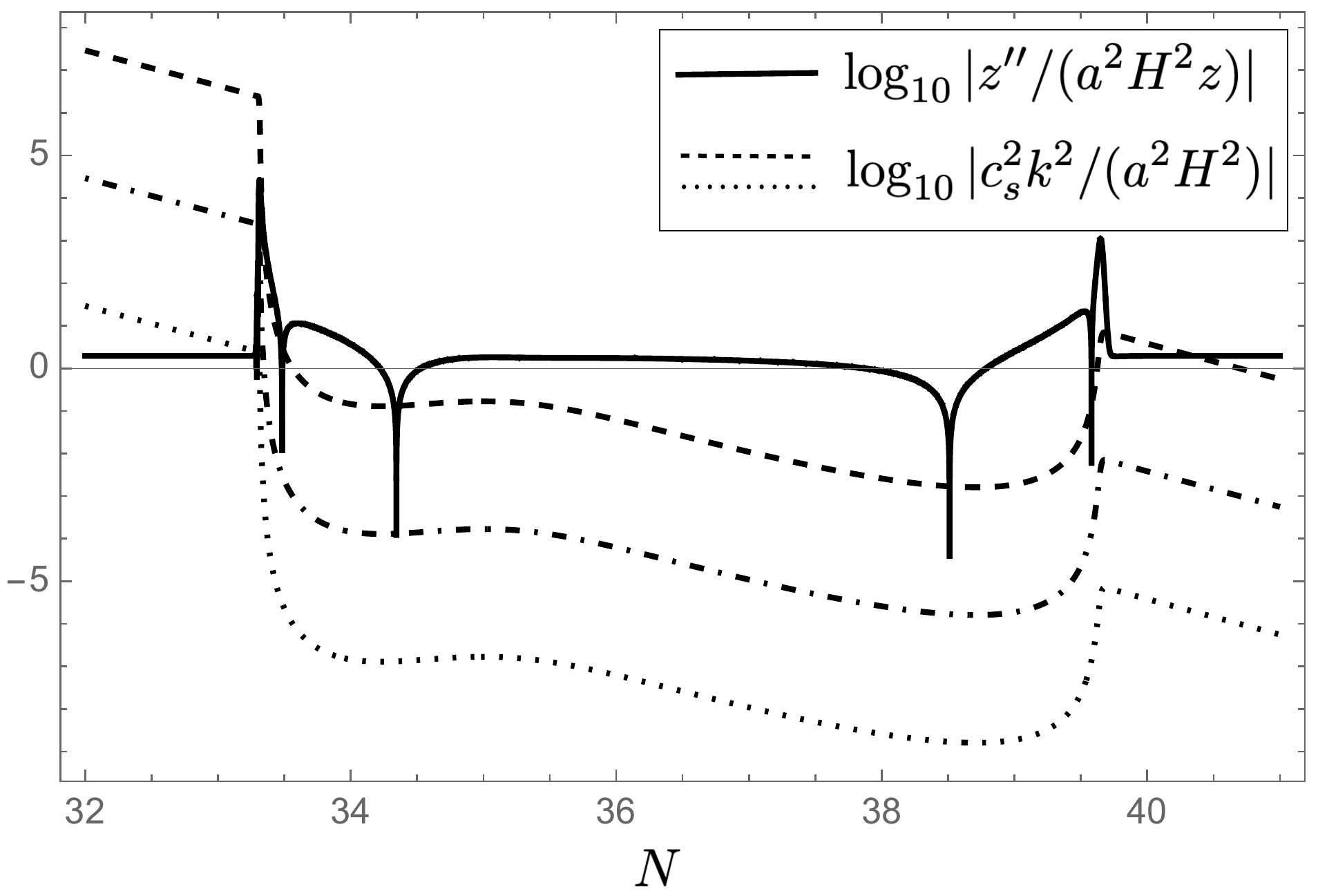}
\caption{\it In the above figure we plot the logarithm of the two main contributions to the time dependent frequency of Eq.(\ref{MS}) normalised by $a^2H^2$ and for different perturbation modes. The solid line represents the behaviour of $z''/z$. We compared the behaviour of the modes $k^2=10^{26}$ (dashed line),  $k^2=10^{23}$ (dashed-dotted line) and $k^2=10^{20}$ (dotted line). When the solid line is below the line representing the evolution of a given mode, the mode is ``inside'' the horizon. Let us note that, in the figure, only the dashed line re-enters the horizon and evolves for $\sim 1$ e-fold inside the horizon after finally exiting it.
\label{figosc}}
\end{figure}
The amplification realised with the choice of parameters (\ref{numfun}) is just an example of how the presence of a sharp feature in the warp factor may affect the scalar power spectrum at the end of inflation. The example shows that the analytical estimates of the previous sections are quite consistent. A suitable choice of $\alpha$, $F$ and $A$ will certainly lead to an amplification of 6-7 orders of magnitude, eventually resulting in PBHs formation with a mass depending on $\Delta N$ (\ref{DNamp}). We restricted our analysis to the study of a few order of magnitude of amplification because a precise numerical analysis and a much steeper warp factor needed by a much larger amount of amplification would have required too much time and processor power. \\
Let us finally note that our results can be easily generalised to different potentials and warp factors. In particular, the steepness required to obtain a large amplification in a short amount of time needs a very localised feature and is therefore very mildly sensitive to the global form of $f$ and $V$.
{ Lastly we observe that models with a varying speed of sound has been discussed in the literature since in the regime with $c_s\ll 1$ a significant amount of non-gaussianities can be generated \cite{varcs}. The common estimates for the abundance of PBHs generated as a consequence of the amplification of the curvature perturbations during inflation assume a gaussian distributed amplitude for such perturbations. The existence of possibly large deviations from gaussianity has been very debated in the literature \cite{nongauss} and, although no general consensus has been found, it is still widely believed that non-gaussianities may improve the efficiency of PBHs formation with a much smaller amount of perturbations amplification.\\
Furthermore, the third order contributions to the action for the perturbations $\mR(x)$ include interactions with couplings which are inverse proportional to $c_s^2$ and may therefore be non-perturbative when $\mR(x)$ is large enough and $c_s^2$ is small. This fact is relevant in particular if one considers effective theories of inflation, defined below a given energy scale, and whose quantum corrections explode for $c_s^2$ below a certain threshold. Let us note, that such terms are multiplied by time derivatives of $\mR$ and, for modes which freeze after the horizon exit, the contribution are therefore suppressed, in contrast with the cases where the amplification is generated by a non-negligible $\dot\mR$ as a consequence of a negative friction term.\\
This question is certainly important but its solution is far from being obvious due to the consequences of non-gaussianities themselves and other details involved in the collapse which are often glossed over but significantly affect the efficiency of PBHs formation. Let us stress, however, that this article only discusses the details of a mechanism for amplification of curvature perturbations in the context of DBI inflation which is supposed to be valid within the linear regime and neglecting the cubic/higher order interactions. If the validity of the linear approximation do not allow for an amplification large enough to generate $100\%$ of PBHs DM, still one may consider the consequences of a smaller amplification phase, leading to a smaller, but significant, amount of PBHs.}
\begin{figure}[t!]
\centering
\includegraphics[width=9.5cm]{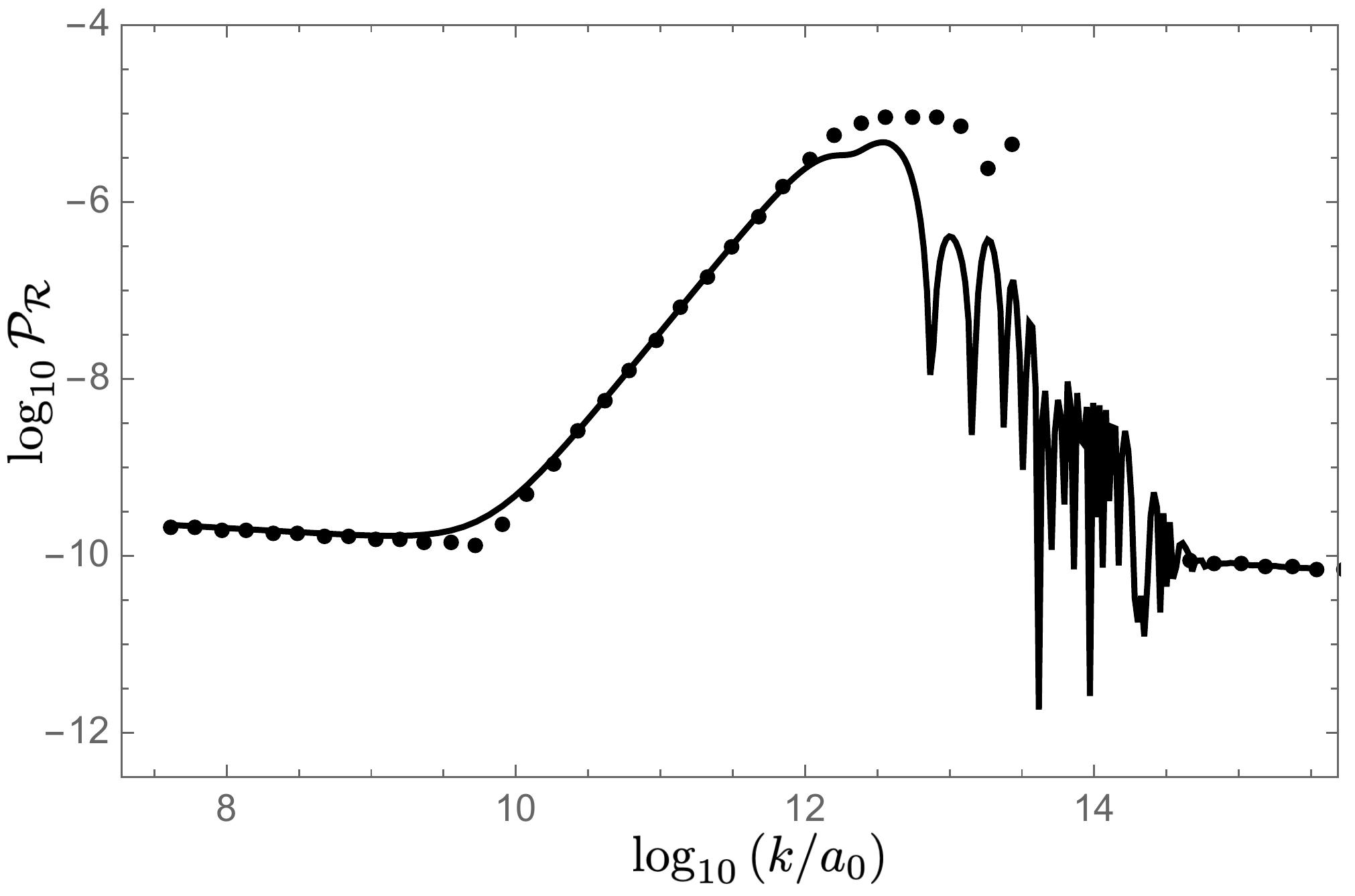}
\caption{\it In the above figure $\log_{10}{\mathcal P}_\mR$ is plotted as a function of $\log_{10}(k/a_0)$. The solid line represents the power spectrum calculated with the exact numerical solution of the Mukhanov-Sasaki equation (\ref{MS}) and the points are the corresponding analytical estimate using Eq. (\ref{PSnoSR}) where the horizon exit is calculated numerically for each mode with the exact (homogeneous) solutions. The points represent the modes which essentially exit the horizon once during their evolution (and some limiting cases for modes which re-enter for a very short amount of time). The oscillation are produced by modes which re-enter the horizon and have enough time to evolve inside the horizon and generate the oscillations before exiting the horizon again. For such modes the estimate (\ref{PSnoSR}) fails and the corresponding points are not plotted. 
\label{fig2}}
\end{figure}
\section{Conclusions}
Primordial black holes have attracted much attention in the last few years since they are viable candidates for dark matter and because the dynamical mechanism behind their formation may probe a range of inflationary scales which are distinct compared to those imprinted in the CMB.\\
In the present article we generalise the conclusions of a preceding paper \cite{ktvv} wherein a mechanism for the amplification of the scalar power spectrum, possibly leading to PBHs formation at the end of inflation, was proposed for a class of non-canonical inflationary models. In such models the inflaton has a non-standard kinetic term and, as a consequence, the speed of sound associated with the curvature perturbations produced during inflation is non-constant and different from the speed of light. In such a context, for a series of simplified models we illustrated that the presence of a decreasing speed of sound may amplify the scalar inflationary perturbations to a critical threshold required in order to lead to an efficient PBHs formation. \\
In particular a string inspired model, wherein a DBI scalar field with a potential and minimally coupled to gravity plays the role of the inflaton, is considered and the possibile amplification of inflationary perturbations due to a decreasing speed of sound is studied.\\
Let us note that, analogously to what occurs for Ultra Slow Roll inflation and a fast variation of $\ep{1}$, also in the presence of a rapidly increasing speed of sound the curvature perturbations may significantly grow after the horizon exit. The corresponding mechanism of amplification has been studied in the literature and applied to DBI inflation \cite{tasinato} but is different from ours which is based on a decreasing speed of sound.\\
For simplicity our results were applied to an inflaton with an exponential potential and warp factor $f(\phi)$ everywhere constant except for a tiny interval where a steep gaussian feature is present. The presence of such a feature leads to a transient phase with a rapidly decreasing speed of sound and a fast amplification of curvature perturbations. We then studied how the homogeneous evolution of the system can be approximated both during SR and in the presence of a rapidly varying warp factor and we found precise analytical expressions for the spectrum of the perturbations generated during these phases. Finally, our approximations were applied to an explicit model where the comparison between the resulting analytical expressions and their exact numerical estimate showed a very good agreement.\\
We finally concluded that DBI inflation is a viable candidate for inflation and may lead to significant PBHs formation in the presence of some very sharp feature in the warp factor. Different DBI inflationary models can be studied as a source of PBHs by means of the precise analytical expressions we found and compared to observational constraints related to inflation and DM abundance. {Non-gaussianities and the validity of the linear approximation in the presence of a small speed of sound remains an open issue. Such a problem must certainly be addressed but it deserves a separate analysis and we must leave it for future research. We however note that the leading contributions to cubic interactions in the $c_s\rightarrow 0$ limit also depend on the time derivative of the curvature perturbations and are therefore suppressed after they exit the horizon and freeze. In contrast, in inflationary models  wherein the amplification of $\mR$ is a consequence of the inversion of the sign of the friction term in the perturbations equation, non-gaussianities become relevant early in the $c_s\rightarrow 0$ limit since the above suppression does not occur.}

\section{Acknowledgements}
A.K. is partially supported by the Russian Foundation for Basic Research grant No. 20-02-00411.


\end{document}